\documentstyle[prd,aps]{revtex}
\begin{document}
\draft
\title{Nonequilibrium dynamics:\\ a renormalized computation scheme}
\author{J\"urgen Baacke
\footnote{Electronic address:~uph001@zx2.hrz.uni-dortmund.de},
Katrin Heitmann \footnote{
Electronic address:~heitmann@hal1.physik.uni-dortmund.de},
Carsten P\"atzold \footnote{
Electronic address:~paetzold@hal1.physik.uni-dortmund.de}
}
\address{Institut f\"ur Physik, Universit\"at Dortmund\\
         D - 44221 Dortmund , Germany}
\date{August 1, 1996}
\maketitle
\begin{abstract}
We present a regularized and renormalized version of the
one-loop nonlinear relaxation equations that determine the
non-equilibrium time evolution of a
classical (constant) field coupled to
its quantum fluctuations. We obtain a computational method
in which the evaluation of divergent fluctuation integrals and the
evaluation
of the exact finite parts are cleanly separated so as to allow
for a wide freedom in the choice of
regularization and renormalization schemes.
We use dimensional regularization here. Within the same formalism
we analyze also the regularization and renormalization of
the energy-momentum tensor. The energy density serves to
monitor the reliability of our numerical computation. 
The method is applied to
the simple case of a scalar $\phi^4$ theory; the results are similar
to the ones found previously by other groups.
\end{abstract}
\pacs{12.15.-y , 11.27.+d , 11.15.Kc, 98.80.Cq }
\section{Introduction} \label{intro}
Phase transitions in elementary particle physics
have been studied over the
last two decades in various contexts. The phase transitions 
of grand unified theories may lead to inflationary
periods of the early universe
\cite{AbPi}; the electroweak phase transition 
can be based on a theory whose parameters - up to the Higgs sector -
are meanwhile well known and studies
both in perturbation theory \cite{BuFoHeWa,FaKaRuSha} and on the
lattice \cite{FaKaLaRuSha,Jan} show considerable progress
in its understanding.
The hadronic phase transition is being investigated both
theoretically and experimentally \cite{Hadro}.

For some applications it is sufficient to work in
finite temperature quantum field theory, i. e. to consider
only states of thermal equilibrium. For the inflationary 
phase transition, at least, a real time study involving nonequilibrium
dynamics is necessary if one intends to describe the
process of reheating. 
The simplest version of the inflationary scenario
which consists in introducing a friction term \cite{AlSt,AbFaWi}
has been shown recently \cite{BVHLS}
to be on weak theoretical grounds. Linear and nonlinear relaxation have
therefore been studied by various authors \cite{BVHLS,BAVHL,Son}. 
The one-loop
relaxation equations studied here do not seem to lead to a
proper thermalization, presumably they have to be modified in
such a way as to include the interactions among the fluctuating
fields. However, they will certainly describe an initial stage
of a more involved dynamical development. They present
therefore a first step that has to be studied and 
well understood. Another subject to which nonequilibrium dynamics
has been applied recently is the hadronic phase transition
\cite{KESCM}
and the possible formation of disordered
chiral condensates \cite{BoVeHo,CKMP}.

The computations of the nonlinear relaxation equations requires
regularization and renormalization. Thus far 
\cite{BVHLS,CKMP} noncovariant
momentum cutoffs were used in such computations. While it would
certainly not be an essential difficulty to replace those by
a Pauli-Villars type regularization one may wish, in computations
involving nonabelian gauge theories, to be able to use dimensional
regularization as well. Furthermore it is more convenient to compute
only convergent integrals instead of dealing
with divergent integrals by varying their cutoff. 

It is the aim of this work to present a computational method in 
which the process of regularization and renormalization is
cleanly separated from the numerical computation of finite
parts of the one-loop integrals. The method is similar to the one 
developed in \cite{Baa1} which has been
applied for the computation of
one loop contributions to reaction rates associated
with sphalerons \cite{BaaJu}, instantons \cite{BaaDai} and
bounces \cite{Baa2}. While in the cases mentioned the computations 
involved Euclidean Green functions which are behaved smoothly,
here we will have to compute the trace of a Green function
in Minkowski space,
i. e. involving real time. The oscillating behaviour
of such Green functions could present a new difficulty for the
application of the computational scheme. 

In this first presentation of the method we have studied its 
practical numerical application only in the simplest possible
model, a scalar field theory with a $\lambda \Phi^4$ self interaction.
This theory has also been the subject of previous studies
\cite{BVHLS} and the results are similar. Especially it is found
that the parametric resonance implied by an oscillating
background field does not fully develop; back reaction damps the
classical amplitude and the system relaxes to a new, essentially
stationary oscillating phase.

The plan of this work is as follows:
in section \ref{basics} we recall the basic definitions and relations;
in section \ref{equ_mot}
 we present the one-loop nonlinear relaxation equations;
we prepare the regularization in section 
\ref{pertex} by expanding the
fluctuation modes in orders of the vertex function governed by the
classical field and by deriving the large momentum behaviour
of the first terms; regularization is then straightforward, 
the renormalization requires some algebra, both are presented in
section \ref{renorm}; the formalism is extended
, in section \ref{EMT}, 
to the renormalization of the energy-momentum tensor;
the numerical computation is discussed 
in section \ref{num_an};
we conclude in section \ref{conclus}
with a discussion of the numerical results
and an outlook to more realistic and more general applications of
the method.  

\section{Basic relations} \label{basics}
 We restrict our study
 to self-interacting scalar $\phi^4$-theory without
spontaneous symmetry breaking. The Lagrangean density is given by
\begin{equation}
{\cal L}=\frac{1}{2}\partial_\mu\Phi\partial^\mu\Phi
-\frac{1}{2}m^2\Phi^2-
\frac{\lambda}{4!}\Phi^4.
\end{equation}
We split the field
$\Phi$ into its expectation value $\phi$ and the quantum fluctuations
$\psi$:
\begin{equation}
\label{erw}
\Phi(x,t)=\phi(t)+\psi(\vec x,t)
\end{equation}
with
\begin{equation}
\phi(t)=\langle\Phi(\vec x,t)\rangle=
\frac{{\rm Tr}{\Phi\rho(t)}}{{\rm Tr}\rho(t)}
\end{equation}
where $\rho(t)$ is the density matrix of the system which satifies
the Liouville equation
\begin{equation}
i \frac{d\rho (t)}{dt} =  [{\cal H}(t),\rho(t)] \; .
\end{equation}
The Lagrangean then takes the form
\begin{equation}
{\cal L}={\cal L}_{\rm 0}+{\cal L}_{\rm I}
\end{equation}
with
\begin{eqnarray}
{\cal L}_{\rm 0}&=&\frac{1}{2}
\partial_\mu\psi\partial^\mu\psi-\frac{1}{2}m^2\psi^2\nonumber\\
&&+\frac{1}{2}
\partial_\mu\phi\partial^\mu\phi-\frac{1}{2}m^2\phi^2-
\frac{\lambda}{4!}\phi^4 \; ,\\
{\cal L}_{\rm I}&=&
\partial_\mu\psi\partial^\mu\phi-m^2\psi\phi-
\frac{\lambda}{4!}\psi^4-\frac{\lambda}{6}\psi^3\phi
-\frac{\lambda}{4}\psi^2\phi^2-\frac{\lambda}{6}\psi\phi^3 \; .
\end{eqnarray}
This decomposition will be used as the basis for a perturbative
expansion. Though we will consider the so-called one-loop equations
which are nonperturbative, we will need such a perturbative scheme
in order to define the one-loop summation and its renormalization.
Since we are dealing with a nonequilibrium system
the appropriate expansion is the CTP formalism
(see e.g. \cite{CalHu}). We will consider the zero temperature case
only.

The Green function is defined by a $2  \times 2$ matrix
\begin{equation}
{\bf G} (t,\vec x;t',\vec x')
= \left\{
\begin{array}{ll} G^{++}(t,\vec x;t',\vec x') &
G^{+-}(t,\vec x;t',\vec x')\\G^{-+}(t,\vec x;t',\vec x') &
G^{--}(t,\vec x;t',\vec x')\end{array} \right\}
\end{equation}
where
\begin{eqnarray}
-iG^{++}(t,\vec x;t',\vec x')&=
&\langle T\Phi(t,\vec x)\Phi(t,\vec x)\rangle\\
-iG^{--}(t,\vec x;t',\vec x')&=&\langle \widetilde{T}\Phi(t,\vec x)
\Phi(t',\vec x')\rangle\\
iG^{+-}(t,\vec x;t', \vec x')&=
&\langle \Phi(t,\vec x)\Phi(t',\vec x')\rangle\\
iG^{-+}(t,\vec x;t',\vec x')&=&iG^{+-}(t',\vec x';t,\vec x) \; .
\end{eqnarray}
Here the brackets $ \langle \rangle $  refer to an expectation
value with respect to the density matrix $\rho$.

The Green functions can be decomposed as
\begin{eqnarray}
\label{green1}
G^{++}(t,\vec x;t',\vec x')&=&G^>(t,\vec x;t',\vec x')
\Theta(t-t')+G^<(t,\vec x;t',\vec x')\Theta(t'-t)\\
\label{green2}
G^{--}(t,\vec x;t',\vec x')&=&G^>(t,\vec x;t',\vec x')
\Theta(t'-t)+G^<(t,\vec x;t',\vec x')\Theta(t-t')\\
\label{green3}
G^{+-}(t,\vec x;t',\vec x')&=&-G^<(t,\vec x;t',\vec x')\\
\label{green4}
G^{-+}(t,\vec x;t',\vec x')&=&-G^>(t,\vec x;t',\vec x')
=-G^<(t',\vec x';t,\vec x)
\end{eqnarray}
with
\begin{equation}
G^>(t,\vec x;t',\vec x')=
i\langle\Phi (t,\vec x)\Phi(t',\vec x')\rangle \; .
\end{equation}

These equations apply to the exact Green functions. For the free
Green functions  
$G^>(t,\vec x;t',\vec x')$ is given explicitly by
\begin{equation}
G^>_0(t,\vec x;t',\vec x')=
\int\!\frac{{\rm d^3}k}{(2\pi)^3}\, \frac{i}{2\omega_k}
\exp(-i \omega_k(t-t')+i \vec k (\vec x-\vec x'))
\end{equation}
with $\omega_k=\sqrt{\vec k^2+m^2}$.

The vertices are obtained from
${\cal L}_{\rm I}$ ordering the terms in powers
of $\psi$
\begin{equation}
{\cal L}_{\rm I}=(\partial_\mu\partial^\mu\phi-
m^2\phi-\frac{\lambda}{6}\phi^3)\psi-\frac{\lambda}{4}\psi^2\phi^2
-\frac{\lambda}{6}\psi^3\phi-\frac{\lambda}{4!}\psi^4 \; .
\end{equation}
Every term corresponds to a vertex. In the matrix form of the
CTP formalism the vertex operators are given by
\begin{equation}
i{\bf \Gamma}_n(t,\vec x)=
\left\{\begin{array}{cc} -i \Gamma_n(t,\vec x)&0 \\
0 & i \Gamma_n(t,\vec x)\end{array} \right\} \; .
\end{equation}
The subscript $n$ denotes the number of fluctuation fields
$\psi$ entering the vertex. We have
\begin{eqnarray}
i\Gamma_1&=&i(\Box-m^2-\frac{\lambda}{6}\phi^2(t))\phi(t) \\
i\Gamma_2&=&-i\frac{\lambda}{2}\phi^2 \\
i\Gamma_3&=&-i\lambda\phi \\
i\Gamma_4&=&-i\lambda \; .
\end{eqnarray}
\section{Equations of motion}
\label{equ_mot}
The equation of motion for the field $\phi(t)$
follows from the tadpole condition \cite{We}
\begin{equation}
\langle\psi^\pm(x,t)\rangle=0 \; .
\end{equation}
After a straightforward calculation one obtains \cite{BVHLS}
the equation 
\begin{equation} \label{phidgl}
\ddot{\phi}(t)+m^2\phi(t)+\frac{\lambda}{6}\phi^3(t)
+\frac{1}{i}\frac{\lambda}{2}\phi(t)G^{++}(0)=0
\end{equation}
which is represented graphically in 
Fig.\ \ref{Fig1}.
Here $G^{++}$ is the $++$ matrix element of 
the exact Green function in the background field
$\phi(t)$. It can be expanded perturbatively as 
\begin{eqnarray}
{\bf G}& =& {\bf G}_0-{\bf G}_0 {\bf \Gamma}_2 {\bf G}_0+ {\bf G}_0
{\bf \Gamma}_2 {\bf G}_0 {\bf \Gamma}_2 {\bf G}_0
-...\nonumber\\
&=&{\bf G}_0\left(1+{\bf\Gamma}_2{\bf G}_0\right)^{-1}\nonumber \; .
\end{eqnarray}
$G^{++}(t,\vec x)$ satisfies the differential equation:
\begin{equation}
\label{dgl}
\left(\frac{\rm \partial^2}{\rm \partial t^2}-
\Delta+m^2+\frac{\lambda}{2}\phi^2(t)\right)
G^{++}(x;x')=\delta^{(4)}(x-x') \; .
\end{equation}
So we have  effectively a time-dependent mass term
\begin{equation}
m^2(t)=m^2+\frac{\lambda}{2}\phi^2(t) \; .
\end{equation}
Using translation invariance in $\vec x$
we introduce the Fourier transform
\begin{equation}
G^{++}_k(t,t')= \int {\rm d}^3x ~ G^{++}(t,\vec x;t',0)
\exp(-i \vec k \vec x)
\end{equation}
and denote the associated energy as
\begin{equation}
\omega_k(t)=\left[ \vec k^2+m^2+
\frac{\lambda}{2}\phi^2(t)\right]^{\frac{1}{2}} \; .
\end{equation}
We make the ansatz  for $G^{++}_k(t,t')$ 
\begin{equation}
G_k^{++}(t,t')=C^{-1}\left\{ U^+_k(t)U^-_k(t')\theta(t-t')
+U^+_k(t')U^-_k(t)\theta(t'-t)\right\} 
\end{equation}
where the $U_k^{\pm}$ are solutions of the homogenous problem and 
$C^{-1}$ is the Wronskian.
The initial conditions are
\begin{eqnarray} \label{incon}
U^+_k(0)&=&1 \hspace{1cm} \dot U^+_k(0)=-i\omega_k^0 \nonumber\\
U^-_k(0)&=&1 \hspace{1cm} \dot U^-_k(0)=i\omega_k^0 
\end{eqnarray}
with
\begin{equation}
\omega_k^0=\left[ \vec {k}^2+m^2+\frac{\lambda}{2}
\phi^2(0)\right]^\frac{1}{2} \; .
\end{equation}
The complete Green function at equal times reads then
\begin{equation}
G_k^{++}(t,t)=\frac{i}{2\omega^0_k}|U^+_k(t)|^2,
\end{equation}
where we have used  $U^-_k(t)=U^{+}_k(t)^*$.

The resulting equation of motion for the classical field
$\phi(t)$ and the mode functions $U_k^+(t)$ are then
\begin{eqnarray}
\label{bewegglg}
\ddot{\phi}(t)+m^2\phi(t)+\frac{\lambda}{6}
\phi^3(t)+\frac{\lambda}{2}\phi(t)\int\!\frac{{\rm d^3}k}{(2\pi)^3}\,
\frac{|U_k^+(t)|^2}{2\omega_k^0}&=&0\\
\label{u+}
\left[\frac{\rm d^2}{\rm dt^2}+
\vec k^2+m^2+\frac{\lambda}{2}\phi^2(t)\right]
U^+_k(t)&=&0\\
U^+_k(0)=1 \hspace{1cm} \dot{U}^+_k(0)&=&-i\omega^0_k \; .
\end{eqnarray}
We denote the fluctuation integral in (\ref{bewegglg}) as
\begin{equation}\label{dmsq}
\Delta {\cal M}^2 (t) 
\equiv \frac{\lambda}{2}\int\!
\frac{{\rm d^3}k}{(2\pi)^3}\,\frac{|U^+_k(t)|^2}{2 \omega_k^0} \; .
\end{equation}
It determines the back-reaction of the fluctuations onto the
classical field $\phi(t)$.

An important check for the consistency of our numerical analysis 
will be the conservation of energy.
The energy density is given by
\begin{equation}
\label{energie}
{\cal E}=\frac{1}{2}\dot{\phi}^2(t)+V(\phi(t))+\frac{{\rm Tr}
{\cal H}\rho(0)}{{\rm Tr} \rho(0)} \; .
\end{equation}
Calculating the trace over the Hamiltonian we obtain
\begin{eqnarray} \label{E_unren}
{\cal E}&=&\frac{1}{2}\dot{\phi}^2(t)+
\frac{1}{2}m^2\phi^2(t)+\frac{\lambda}{4!}\phi^4(t) \nonumber \\
&&+\int\!\frac{{\rm d^3}k}{(2\pi)^3}\,
\frac{1}{2\omega_k^0}\left\{\frac{1}{2}|\dot{U}_k^+(t)|^2+\frac{\vec
k^2}{2}|U_k^+(t)|^2+\frac{1}{2}m^2(t)|U_k^+(t)|^2\right\} \; .
\end{eqnarray}
Using the equations of motion it is easy to see that
the time derivative of the energy density vanishes.
The equations obtained so far are yet formal since they
contain divergent quantities. We will present their renormalized
form in section \ref{renorm}.  

\section{Perturbative expansion} \label{pertex}
In order to prepare the renormalized version of the
equations given in the previous section we introduce a suitable
expansion of the
mode functions $U_k^{\pm}(t)$.
Adding the term $\frac{\lambda}{2}\phi^2(0)U_k^+(t)$
on both sides of the mode function equation it takes 
the form
\begin{equation}
\label{udgl}
\left[ \frac{\rm d^2}{\rm dt^2}+
(\omega^0_k)^2\right]U_k^+(t)=-V(t)U_k^+(t)
\end{equation}
with
\begin{eqnarray}
V(t)&:=&\frac{\lambda}{2}\left(\phi^2(t)-\phi^2(0)\right)
\nonumber\\
\omega_k^0&=&\left[\vec k^2+
m^2+\frac{\lambda}{2}\phi^2(0)\right]^{1/2} \; .
\end{eqnarray}
Including the initial conditions (\ref{incon}) 
the mode functions satisfy the equivalent integral equation
\begin{equation}
U_k^+(t)=e^{-i\omega_k^0 t}+\int\limits^{\infty}_{0}\!{\rm d}t'
\Delta_{k,{\rm ret}}(t-t')V(t')U_k^+(t')
\end{equation}
with
\begin{equation}
\label{fvt}
\Delta_{k,{\rm ret}}(t-t')=
-\frac{1}{\omega_k^0}\Theta(t-t')\sin(\omega_k^0(t-t')) \; .
\end{equation}
We separate $U_k^+(t)$ into the trivial part corresponding to
the case $V(t)=0$ and a function $f_k(t)$ which represents the
reaction
to the potential by making the ansatz
\begin{equation}
\label{ansatz}
U_k^+(t)=e^{-i\omega_k^0 t}(1+f_k(t)) \; .
\end{equation}
$f_k(t)$ satisfies then the integral equation
\begin{equation}\label{finteq}
f_k(t)=\int\limits^{t}_{0}\!{\rm d}t'\Delta_{k,{\rm ret}}(t-t')V(t')
(1+f_k(t'))e^{i\omega_k^0 (t-t')} 
\end{equation}
and an equivalent differential equation
\begin{equation}\label{fdiffeq}
\ddot{f}_k(t)-2i\omega_k^0\dot{f}_k(t)=-V(t)(1+f_k(t)) 
\end{equation}
with the initial conditions $f_k(0)=\dot{f}_k(0)=0$.
\\ \\
We expand now $f_k(t)$ with respect to orders in $V(t)$
by writing
\begin{eqnarray}
\label{entwicklung}
f_k(t)&=& f_k^{(1)}(t)+f_k^{(2)}(t)+f_k^{(3)}(t) + .... \\
 &=& f_k^{(1)}(t)+f_k^{\overline{(2)}}(t)
\end{eqnarray}
where $f_k^{(n)}(t)$ is of n'th order in $V(t)$ and 
$f_k^{\overline{(n)}}(t)$
is the sum over all orders beginning with the n'th one. 
The $f_k^{(n)}$ are obtained by iterating the integral
equation (\ref{finteq}) or the differential equation
(\ref{fdiffeq}). The function $f_k^{\overline{(1)}}(t)$ is
identical to the function $f_k(t)$ itself which is obtained
by solving (\ref{fdiffeq}), the function
$f_k^{\overline{(2)}}(t)$ can be obtained
as
\begin{equation}\label{f2inteq}
f_k^{\overline{(2)}}(t)=
\int\limits^{t}_{0}\!{\rm d}t'\Delta_{k,{\rm ret}}(t-t')V(t')
f_k^{\overline{(1)}}(t')e^{i\omega_k^0 (t-t')} 
\end{equation}
or by solving the inhomogeneous differential equation
\begin{equation}\label{f2diffeq}
\ddot{f}_k^{\overline{(2)}}(t)-2i\omega_k^0\dot{
f}_k^{\overline{(2)}}(t)=-V(t)f_k^{\overline{(1)}}(t) \; .
\end{equation}
Note that in this way one avoids the computation 
of $f_k^{\overline{(2)}}(t)$ 
via the small difference  $f_k(t)-f_k^{(1)}(t)$. This feature is 
especially important if deeper subtractions are required as in the
case of fermion fields. 

The order on the potential $V(t)$
will determine the behaviour of the functions
$f_k^{(n)}$ at large momentum.
We will give here the relevant leading terms for $f_k^{(1)}(t)$ and
$f_k^{(2)}(t)$. We have
\begin{equation}
f_k^{(1)}(t)=\frac{i}{2 \omega_k^0}
\int\limits^{t}_{0}\!{\rm d}t'
(\exp(2 i \omega_k^0(t-t'))-1)V(t')  \; .
\end{equation}
Integrating by parts we obtain
\begin{equation}\label{f1exp}
f_k^{(1)}(t)= -\frac{i}{2\omega_k^0}\int\limits^{t}_{0}\!{\rm d}t'
V(t')-\frac{1}{4(\omega_k^0)^2}V(t)
+\frac{1}{4(\omega_k^0)^2}\int\limits^{t}_{0}\!{\rm d}t'
\exp(2 i \omega_k^0(t-t'))\dot{V}(t')
\end{equation}
or, by another integration by parts
\begin{eqnarray}
f_k^{(1)}(t)&=& -\frac{i}{2\omega_k^0}\int\limits^{t}_{0}\!{\rm d}t'
V(t')-\frac{1}
{4(\omega_k^0)^2}V(t)+\frac{i}{8(\omega_k^0)^3} \dot{V}(t) \\
&&-\frac{i}{8(\omega_k^0)^3}\int\limits^{t}_{0}\!{\rm d}t'
\exp(2 i \omega_k^0 (t-t'))\ddot{V}(t') \; .
\end{eqnarray}
Similarly we find for the leading behaviour of $f_k^{(2)}(t)$
\begin{equation}\label{f2exp}
f_k^{(2)}(t)= -\frac{1}{4(\omega_k^0)^2}
\int\limits^{t}_{0}\!{\rm d}t'\int\limits^{t'}_{0}\!{\rm d}t''
V(t')V(t'') + O((\omega_k^0)^{-3}) \; .
\end{equation}
The leading terms of 
$f_k^{\overline{(1)}}(t)$ and $f_k^{\overline{(2)}}(t)$
in this expansion in powers of $(\omega_k^0)^{-1}$ are the same
as for $f_k^{(1)}(t)$ and $f_k^{(2)}(t)$ respectively.

Unlike a WKB ansatz for $U_k^+(t)$ the expansion presented here can
be easily extendend \cite{Baa1} 
to coupled channel systems and higher orders in the
expansion (if deeper subtractions are required).

\section{Renormalization} \label{renorm}
The fluctuation term (\ref{dmsq}) occuring in the 
equation of motion
(\ref{bewegglg}) can then be written as
\begin{equation}
\label{u}
\Delta {\cal M}^2 (t) 
=\frac{\lambda}{2}\int\!
\frac{{\rm d^3}k}{(2\pi)^3}\,
\frac{|e^{-i\omega_k^0 t}(1+f_k(t))|^2}{2 \omega_k^0} \; .
\end{equation}
Inserting our expansion we obtain
\begin{eqnarray}
\label{erg}
\Delta {\cal M}^2 (t)
&=& 
 \frac{\lambda}{2}\int\!\frac{{\rm d^3}k}{(2\pi)^3}\,
\frac{1}{2 \omega_k^0}\left\{1+2 {\rm Re}f_k^{(1)}(t)\right\}
\\ \nonumber
&&+\frac{\lambda}{2}\int\!
\frac{{\rm d^3}k}{(2\pi)^3}\,\frac{1}{2 \omega_k^0}
\left\{2 {\rm Re}f_k^{\overline 
{(2)}}(t) 
+|f_k^{\overline {(1)}}(t)|^2\right\} \; .
\end{eqnarray}

The second integral is convergent. Though 
${\rm Re}f_k^{\overline{(2)}}(t)$ and $|f_k^{\overline {(1)}}(t)|^2$
behave separately only as $(\omega_k^0)^{-2}$ one finds, using
Eqs. (\ref{f1exp}) and (\ref{f2exp}), that
this leading behaviour cancels
among the two contributions. This integral can be computed numerically
using the functions $f_k^{\overline{(1)}}$ and
$f_k^{\overline{(2)}}$ obtained by solving (\ref{fdiffeq}) and
(\ref{f2diffeq}). 
In the first integral we find a quadratic 
divergence which corresponds to the tadpole graph in $\phi^4$-theory
and a logarithmic one associated with $ 2 {\rm Re}f_k^{(1)}(t)$.

Indeed the real part of $f_k^{(1)}(t)$ is 
obtained from (\ref{f1exp}) as
\begin{equation}
2{\rm Re}f_k^{(1)}(t)
=-\frac{1}{2{\omega_k^0}^2}\left\{V(t)-\int\limits^{t}_{0}\!
{\rm d}t'\cos(2\omega_k^0(t-t'))\dot{V}(t')\right\} \; .
\end{equation}
The first term behaves as $(\omega_k^0)^{-2}$ and leads therefore
to a logarithmic divergence in $\Delta {\cal M}^2$. The second one
yields a finite contribution. It will be
included into the finite part of $\Delta {\cal M}^2$.

So altogether we have
\begin{equation}
\label{skalare div}
\Delta {\cal M}^2(t) =
\frac{\lambda}{2}
\int\!\frac{{\rm d^3}k}{(2\pi)^3}\,\frac{1}{2 \omega_k^0}-
\frac{\lambda}{2}
\int\!\frac{{\rm d^3}k}{(2\pi)^3}\,
\frac{1}{2\omega_k^0}\frac{V(t)}{2(\omega_k^0)^2}
+\Delta {\cal M}^2_{{\rm fin}}(t)
\end{equation}
with
\begin{eqnarray}
\label{numerik}
\Delta {\cal M}^2_{{\rm fin}}(t)
 &=&\frac{\lambda}{2}\int\!\frac{{\rm d^3}k}{(2\pi)^3}\,\frac{1}{2
\omega_k^0}\frac{1}{2(\omega_k^0)^2}\int\limits^{t}_{0}\!
{\rm d}t'\cos(2\omega_k^0(t-t'))\dot{V}(t')\nonumber\\
&+&\frac{\lambda}{2}
\int\!\frac{{\rm d^3}k}{(2\pi)^3}\,\frac{1}{2 \omega_k^0}\left\{2 {\rm
Re}f_k^{\overline {(2)}}(t)
+|f_k^{\overline {(1)}}(t)|^2\right\} \; .
\end{eqnarray}
The first term in (\ref{skalare div}) corresponds to the zeroth 
order diagram in Fig.\ \ref{Fig1}. The second term in (\ref{skalare div})
and the first term in (\ref{numerik}) together correspond to the first
order diagram in Fig.\ \ref{Fig1}, the second term in (\ref{numerik})
sums up all the higher order diagrams. 
By this term-to-term equivalence of our expressions
to diagrams of CTP perturbation theory 
we have achieved here a clean separation between
finite quantities that
can be computed numerically and the renormalization parts that
are computed analytically up to Fourier transforms of the
external sources. 
We consider this to be an important feature of our method, as
a the regularization may now be chosen freely, as
required in order to maintain the symmetries of the theory.
 
The first divergent integral can be rewritten as
\begin{equation}\label{tadpole}
\int\!\frac{{\rm d^3}k}
{(2\pi)^3}\,\frac{1}{2 \omega_k^0}=\int\frac{{\rm d}^4k}{(2\pi)^4}
\frac{i}{k^2-m^2+i o} \; ,
\end{equation}
an identity that holds in dimensional as well as in Pauli-Villars
regularization. The integral is the one associated to the tadpole graph
Fig.\ \ref{Fig2}a and can be absorbed into the renormalization of the mass
term.  In the same way the second integral is equivalent to
\begin{equation}\label{fish}
\int\!\frac{{\rm d^3}k}{(2\pi)^3}\,\frac{1}{8(\omega_k^0)^3}=
\frac{1}{2}\int\frac{{\rm d}^4 k}{(2\pi)^4}\frac{-i}{(k^2-m^2+i o)^2}
\; .
\end{equation}
It is associated with the Feynman graph Fig.\ \ref{Fig2}b which leads to a
coupling constant renormalization.
We will use dimensional regularization and renormalize the
four point function at vanishing external momenta.
We first rewrite
the basic equation of motion, including appropriate counter terms, as
\begin{equation}
\ddot{\phi}(t)+(m^2+\delta m^2)\phi(t)+
\frac{\lambda+\delta \lambda}{6}\mu^{\epsilon}\phi^3(t)+
\Delta {\cal M}^2(t)\phi(t)=0 \; .
\end{equation}
Next we separate from $\Delta {\cal M}^2(t) \phi(t)$
dimensionally regularized divergent terms
\begin{eqnarray}
\label{dimreg1}
\left\{\frac{\lambda}{2}\phi(t)
\int\!\frac{{\rm d^3}k}{(2\pi)^3}\, \frac{1}{2\omega_k^0}\right\}_
{{\rm reg}}
&=&\mu^\epsilon\frac{\lambda}{2}\phi(t)\int
\!\frac{{\rm d^d} k}{(2\pi)^d}\,
\frac{1}{2\left[\vec k^2+m^2+\frac{\lambda}{2}\mu^\epsilon
\phi^2(0)\right]^\frac{1}{2}}
\nonumber\\
&=&-\frac{\lambda m_0^2 
\phi(t)}{32\pi^2}\left\{\frac{2}{\epsilon}+
\ln{\frac{4\pi\mu^2}{m_0^2}}-\gamma+1\right\}
\end{eqnarray}
and
\begin{eqnarray}
\label{dimreg2}
\left\{-\frac{\lambda}{2}\phi(t)
V(t)\int\!\frac{{\rm d^3}k}{(2\pi)^3}\, \frac{1}{4{\omega_k^0}^3}
\right\}_{{\rm reg}}&=&-\frac{1}{8}(\mu^2)
^\epsilon\lambda\phi(t) V(t)\int\!
\frac{{\rm d^d} k}{(2\pi)^d}\,
\frac{1}{\left[\vec k^2+m^2+
\frac{\lambda}{2}\mu^\epsilon\phi^2(0)\right]^\frac{3}{2}}
\nonumber\\
&=& -\frac{\lambda \mu^\epsilon \phi(t)V(t)}
{32\pi^2}\left\{\frac{2}{\epsilon}
+\ln{\frac{4\pi\mu^2}{m_0^2}}-\gamma\right\}
\end{eqnarray}
where we have introduced
\begin{equation} \label{inimass}
m_0^2:=m^2+\frac{\lambda}{2}\mu^\epsilon\phi^2(0)   \; .
\end{equation}
Recalling that $V(t)=(\lambda/2)(\phi^2(t)-\phi^2(0))$ these two
expressions combine into
\begin{equation} \label{combin}
-\delta m^2 \phi(t) - \frac{\delta \lambda \mu^{\epsilon}}{6}
\phi^3(t) -\frac{\lambda}{32 \pi^2}\ln\frac{m^2}{m_0^2}
\left(m^2\phi(t)+\frac{\lambda}{2}\phi^3(t)\right)-
\frac{\lambda^2}{64\pi^2}\mu^\epsilon\phi^2(0)\phi(t)
\end{equation}
where we have defined the renormalization constants
\begin{equation}                 \label{deltam}
\delta m^2 =
\frac{\lambda m^2}{32\pi^2}\left\{\frac{2}{\epsilon}+
\ln{\frac{4\pi\mu^2}{m^2}}-\gamma+1\right\}
\end{equation}
and
\begin{equation} \label{deltalam}
\delta \lambda =
\frac{3\lambda^2}{32\pi^2}\left\{\frac{2}{\epsilon}+
\ln{\frac{4\pi\mu^2}{m^2}}-\gamma\right\} \; .
\end{equation}
Note that these counterterms are
independent of $m_0$ and therefore from
the initial value of the field $\phi(0)$, though the divergent
integrals do depend on the initial condition. 

The renormalized equation of motion now reads
\begin{equation}
\ddot{\phi}(t)+(m^2+\Delta m^2)\phi(t)+
\frac{\lambda+\Delta \lambda}{6}\phi^3(t)+
\Delta {\cal M}^2_{{\rm fin}}(t)\phi(t)=0
\end{equation}
with the finite corrections
\begin{eqnarray}
\Delta m^2 &=&-\frac{\lambda m^2}{32\pi^2}\ln\frac{m^2}{m_0^2}
-\frac{\lambda^2}{64 \pi^2}\phi^2(0) \; ,
\\
\Delta \lambda&=&-\frac{3\lambda^2}{32\pi^2}\ln\frac{m^2}{m_0^2} \; .
\end{eqnarray}
This equation is in a form well suited for numerical
computation.

\section{The energy-momentum tensor}
\label{EMT}

The renormalization scheme applied so far concerned the
Green functions of the theory. The energy density on the other hand
is part of the energy-momentum tensor. It contains additional
divergent terms which have to be defined by a regularization and
to be removed by new counter terms. These counter terms have been
considered in the literature \cite{CaCoJa,Chr,BoHoVeSa,CoHaKlMo}.
If one considers a space with vanishing curvature
one has only two possible counter terms besides the ones already
introduced into the Lagrangean so that
\begin{equation} \label{T_counterms}
T_{\mu\nu}^{\rm ren}=T_{\mu\nu}^{\rm reg}+ g_{\mu\nu}
\left(\delta\Lambda+\frac{1}{2}\delta m^2 \phi^2(x)
+\frac{\delta \lambda}{4!} \phi^4(x)\right) +
A (g_{\mu\nu}\partial_\alpha\partial^\alpha
-\partial_\mu\partial_\nu)
\phi^2(x) \; .
\end{equation}
In the case considered here the second term does
not contribute to the energy density. It will be useful, however, 
to consider the entire energy-momentum tensor including the space
components
\begin{equation}
T_{ij}=-p g_{ij}
\end{equation}
where $p$ is the pressure
 given formally (i. e. without regularization) by
\begin{equation}
\label{pressure}
p=\dot\phi^2(t)
+\int\!\frac{d^3k}{(2\pi)^3}\,\frac{1}{2\omega_k^0}
\left(|\dot 
U_k^+(t)|^2+\frac{\vec k^2}{3}|U_k^+(t)|^2
\right)-{\cal E}\; .
\end{equation}
The constants $A$ and $\delta \Lambda$ have been determined by various
authors \cite{CaCoJa,Chr,BoHoVeSa,CoHaKlMo}
 using different regularizations.
Coleman and Jackiw use the covariant Pauli-Villars
regularization and Christensen the point-splitting technique
and heat kernel regularization. Cooper et al. find a way of
dealing with the problems of a three-momentum cutoff.
We will use the covariant dimensional
regularization as we did above in regularizing the
fluctuation integral.

In order to renormalize the
energy we introduce the available counter terms 
into the 
unrenormalized expression (\ref{E_unren}) so that it reads now
\begin{eqnarray}\label{E_ren1}
{\cal E}&=&\frac{1}{2}\dot{\phi}^2(t)+
	\frac{1}{2}(m^2+\delta m^2)\phi^2(t)
+\frac{\lambda+\delta\lambda}{4!}\mu^\epsilon\phi^4(t)
+ \delta\Lambda\nonumber\\
&&+\int\!\frac{{\rm d^3}k}{(2\pi)^3}\,
\left\{\frac{\omega_k^0}{4}(1+2{\rm Re}
f_k^{\overline{(1)}}(t)+|f_k^{\overline{(1)}}(t)|^2)\right.\nonumber\\
&&\hspace{2.4cm}+
\frac{1}{4\omega_k^0}|\dot{f}_k^{\overline{(1)}}(t)|^2\nonumber\\
&&\hspace{2.4cm}-
\frac{1}{2}{\rm Re}
(i\dot{f}_k^{\overline{(1)}*}(t)+
if_k^{\overline{(1)}}(t)\dot{f}_k^{\overline{(1)}*}(t))\nonumber\\
&&\hspace{2.4cm}\left.+\frac{1}{4}(\omega_k^0+\frac{V(t)}{\omega_k^0})
(1+2{\rm Re}f_k^{\overline{(1)}}(t)
+|f_k^{\overline{(1)}}(t)|^2)\right\} \; .
\end{eqnarray}

While $\delta \lambda$ and $\delta m^2$ have already
been fixed above we have to determine now the `cosmological constant'
$\delta \Lambda$; in order to do so
 we analyse the quartically divergent part of the 
fluctuation integral
\begin{equation}
{\cal E}_{\rm quart} =\int\!\frac{{\rm d^3}k}{(2\pi)^3}
\,\frac{\omega^0_k}{2} \; .
\end {equation}
In dimensional regularization it becomes
\begin{equation}\label{quart}
{\cal E}_{\rm quart} =-\frac{m^4_0}{64 \pi^2}\left\{\frac{2}
{\epsilon}+\ln \frac{4\pi \mu^2}{m_0^2} +\frac{3}{2}
-\gamma \right \}\; .
\end {equation}
This expression obviously depends  on the initial condition via
the `initial mass' $m_0$. If we insert the expression
(\ref{inimass}) we find
\begin{equation}
{\cal E}_{\rm quart} =-\frac{1}{64 \pi^2}
\left(m^4+\lambda m^2\phi^2 (0)+
\frac{\lambda^2}{4}\phi^4(0)\right)\left\{\frac{2}
{\epsilon}+\ln \frac{4\pi \mu^2}{m_0^2} +\frac{3}{2}
-\gamma \right \} \; .
\end {equation}
The divergent parts proportional 
to $\phi^2(0)$ and $\phi^4(0)$ will be cancelled by divergent
parts of the fluctuation integral (see below). We define here the 
`cosmological constant' counter term by
\begin{equation}
\delta \Lambda =\frac{m^4}{64\pi^2}\left(\frac{2}{\epsilon}+
\ln \frac{4 \pi \mu^2}{m^2}-\gamma+\frac{3}{2}\right) \; , 
\end{equation}
again independent of the initial condition,
and retain in the expression for the energy a finite term
\begin{equation}
\Delta \Lambda =-\frac{m^4}{64\pi^2} \ln \frac{m_0^2}{m^2}
\; .
\end{equation}
Having considered the quartically divergent term
in the fluctuation integral we turn now
to the quadratic ones. One can show that the terms 
$\omega_k^0{\rm Re} f_k^{(1)}(t)/2$ and
$-{\rm Re}(i\dot{f}^{(1)*}(t))/2$ cancel.
The term proportional to $\frac{V(t)}{4\omega_k^0}$
gives in dimensional regularization
\begin{equation}
{\cal E}_{\rm quad}= 
\int\!\frac{{\rm d^3}k}{(2\pi)^3}\frac{V(t)}{4\omega_k^0}
=-\frac{\lambda m_0^2}{64 \pi^2}
\left( \phi(t)^2 - \phi(0)^2\right)
\left\{\frac{2}{\epsilon}+\ln\frac{4\pi\mu^2}{m_0^2} -\gamma +1
\right\}\;.
\end{equation}
These terms have to be considered together with the 
logarithmically divergent ones which are proprotional to
$V(t)^2$. Their explicit form follows from an analysis of the
leading behaviour.
It combines to
\begin{equation}
{\cal E}_{\rm log} =- V(t)^2\int\!\frac{{\rm d^3}k}{(2\pi)^3}
\frac{1}{16 (\omega_k^0)^3}
=-\frac{\lambda^2\mu^\epsilon}{256\pi^2}
\left( \phi(t)^2 - \phi(0)^2\right)^2
\left\{\frac{2}{\epsilon}+\ln\frac{4\pi\mu^2}{m_0^2} -\gamma 
\right\} \;.
\end{equation}
Instead of removing the complete terms
$ |\dot{f}_k^{\overline{(1)}}(t)|^2/4\omega_k^0$ and 
$2 V(t){\rm Re}f_k^{\overline{(1)}}(t)/4\omega_k^0$
from the fluctuation integral we just subtract the
leading behaviour from the integrand, this is algebraically 
less cumbersome at the expense of inducing a small difference of large 
quantities (which was found to be tolerable).
As before (see (\ref{combin})) 
the quadratically and logarithmically divergent
terms combine in such a way that the $m_0$ is replaced by
$m$ in the divergent parts.
The terms proportional to $\phi^2(t)$ and $\phi^4(t)$ are cancelled 
by the mass renormalzation and coupling constant renormalization
counter terms up to finite contributions. The terms
proportional to  $\phi^2(0)$ and $\phi^4(0)$ are cancelled
by the corresponding terms occuring the quartically divergent part
(\ref{quart}), again up to a finite remainder
\begin{equation}
\Delta\Lambda'=\frac{1}{128\pi^2}\left(\frac{\lambda^2}{4}
\phi^4(0)-\lambda m^2 \phi^2(0)\right)\; .
\end{equation} 
We finally obtain for the energy density the renormalized
expression
\begin{eqnarray}
{\cal E}_{\rm ren}&=&\frac{1}{2}\dot{\phi}^2(t)+
\frac{1}{2}(m^2+\Delta m^2)\phi^2(t)
+\frac{\lambda+\Delta \lambda}{4!}\phi^4(t)
+\Delta \Lambda\ +\Delta\Lambda'\nonumber\\
&&+\int\!\frac{{\rm d^3}k}
{(2\pi)^3}\,\left\{\frac{\omega_k^0}{2}(2{\rm Re}
f_k^{\overline{(2)}}(t)+|f_k^{\overline{(1)}}(t)|^2)\right.\nonumber\\
&&\hspace{2.4cm}
+\frac{1}{4\omega_k^0}|\dot{f}_k^{\overline{(1)}}(t)|^2\nonumber\\
&&\hspace{2.4cm}-
\frac{1}{2}{\rm Re}
(i\dot{f}_k^{\overline{(2)}*}(t)+
if_k^{\overline{(1)}}(t)\dot{f}_k^{\overline{(1)}*}(t))\nonumber\\
&&\hspace{2.4cm}\left.+\frac{V(t)}{4\omega_k^0}
(2{\rm Re}f_k^{\overline{(1)}}(t)
+|f_k^{\overline{(1)}}(t)|^2)+\frac{V^2(t)}{16{\omega_k^0}^3}\right\} \; .
\end{eqnarray}
In the expression (\ref{pressure}) for the pressure the
energy ${\cal E}$ has to be replaced by the
renormalized one. This replacement absorbs all the
counter terms proportional to $g_{\mu\nu}$ (see
(\ref{T_counterms})). The remaining counter term 
has to be added and the constant $A$ has to be chosen
so as to cancel the remaining divergencies.
Replacing at the same time the mode functions 
$U_k^+$ by their expansion in terms of the $f_k^{(n)}$
we find
\begin{equation}\label{p_renorm}
p_{\rm ren}=-{\cal E}_{\rm ren}+
\dot\phi^2(t)
+ A \frac{\rm d^2}{{\rm d}t^2}\phi^2(t) +p_{\rm fluct}
\end{equation}
with
\begin{eqnarray}
p_{\rm fluct} =
&&\int\!\frac{d^3k}{(2\pi)^3}\,\frac{1}{2\omega_k^0}
\left\{ \left((\omega_k^0)^2+(\vec k)^2/3\right)
\left(1 +2 {\rm Re} f_k^{\overline{(1)}}(t)
+|f_k^{\overline{(1)}}(t)|^2\right)\nonumber\right. \\
&&\left.+ |\dot{f}_k^{\overline{(1)}}(t)|^2
-2 \omega_k^0 {\rm Re}\left(1+f_k^{\overline{(1)}}(t)\right)
\dot{f}_k^{\overline{(1)}*}(t)\right\} \; .
\end{eqnarray}
This expression looks at first sight hopelessly divergent 
and there is just one counter term to cancel these divergencies.
Using dimensional regularization we find
that the leading quartic divergence reduces to a finite term
via
\begin{eqnarray}
\left\{\int\!\frac{{\rm d^3}k}{(2\pi)^3}\,
\frac{1}{2\omega_k^0}({(\omega_k^0)}^2+\frac{\vec k^2}{3}) \right\}
_{\rm reg}
&=&(\mu)^{\epsilon}\int\!\frac{{\rm d^{d}}k }{(2\pi)^{d}}\,
\frac{1}{2\omega_k^0}({(\omega_k^0)}^2+\frac{\vec k^2}{3})\nonumber\\
&=&-\frac{m_0^4}{96\pi^2}\; .
\end{eqnarray}
The terms leading to a quadratic divergence are
\begin{equation}
\omega_k^0{\rm Re} f^{(1)}(t)-{\rm Re}(i\dot{f}^{(1)*}(t))+\frac{\vec k^2}{3 \omega_k^0}{\rm Re} 
f^{(1)}(t)\; .
\end{equation}
Analyzing their leading behaviour one finds that the possible
quadratic divergence reduces  again to a finite
result via
\begin{eqnarray}
&&\left\{V(t)\int\!\frac{{\rm d^3}k}{(2\pi)^3}\,\frac{1}{6\omega_k^0}+m_0^2 V(t)\int\!\frac{{\rm d^3}k}{(2\pi)^3}\,\frac{1}{12{(\omega_k^0)}^3}
\right\}_{\rm reg}\nonumber\\
&=&-V(t)\frac{m_0^2}{48\pi^2}\left(\frac{2}{\epsilon}
+\ln{\frac{4\pi\mu^2}{m_0^2}}-\gamma+1\right)
+V(t)\frac{m_0^2}{48\pi^2}\left(\frac{2}{\epsilon}
+\ln{\frac{4\pi\mu^2}{m_0^2}}-\gamma\right)\nonumber\\
&=&-\frac{V(t)m_0^2}{48\pi^2} \;.
\end{eqnarray}
After evaluating these leading divergencies we find
\begin{eqnarray}
p_{\rm fluct} &=&-\frac{m_0^4}{96\pi^2}-\frac{V(t)m_0^2}{48\pi^2}+
\nonumber\\
&&\hspace{-1.3cm}+\int\!\frac{{\rm d^3}k}{(2\pi)^3}\,\frac{1}{2\omega_k^0}
\left\{({(\omega_k^0)}^2+\frac{\vec k^2}{3})\left[
\frac{1}{4{(\omega_k^0)}^3}\int\limits_0^{t}\!{\rm d}t\,\sin{2\omega_k^0(t-t')\ddot{V}(t')}
+2{\rm Re}f_k^{\overline{(2)}}(t)+|f_k^{\overline{(1)}}(t)|^2
\right]\right.\nonumber\\
&&\hspace{-1.3cm}+|f_k^{\overline{(1)}}(t)|^2-\frac{1}{2\omega_k^0}\int\limits_0^{t}\!{\rm d}t\,
\sin{2\omega_k^0(t-t')}\ddot{V}(t')\nonumber\\
&&\hspace{-1.3cm}+2{\rm Re}(-i\omega_k^0\dot{f}^{\overline{(1)}}(t)
-i\omega_k^0f_k^{\overline{(1)}}(t)
f_k^{\overline{(1)}*}(t))
\biggr\}\; .
\end{eqnarray}
The analysis of the logarithmically divergent terms becomes now
very cumbersome. After some algebra and integrations
by parts one finds the logarithmically divergent term
\begin{eqnarray}
\label{logdiv}
&&\ddot{V}(t)\left\{\int\!\frac{{\rm 
d^3}k}{(2\pi)^3}\,\frac{1}{2\omega_k^0}\left\{\frac{1}{8{(\omega_k^0)}^4}
\left( {(\omega_k^0)}^2+\frac{\vec k^2}{3} \right)
-\frac{1}{4{(\omega_k^0)}^2} \right\}\right\}_{\rm reg}\nonumber\\
&&=-\frac{\ddot{V}(t)}{96\pi^2}\left\{\mu^{\epsilon}\left(\frac{2}{\epsilon}
+\ln{\frac{4\pi\mu^2}{m_0^2}}-\gamma\right)+\frac 1 3\right\}\; .
\end{eqnarray}
This determines the divergent part of the renormalization constant
$A$. The finite part can be chosen freely and constitutes a free
parameter of the theory, used e.g. for the `improved' 
energy-momentum tensor. We dispose of this freedom by choosing
\begin{equation}
A=
-\frac{\lambda}{192\pi^2}\left\{\mu^{\epsilon}\left(\frac{2}{\epsilon}
+\ln{\frac{4\pi\mu^2}{m_0^2}}-\gamma\right)
\right\}
\end{equation}
since then the expression in parentheses is just the 
standard equivalent of $\ln (\Lambda^2/m^2)$ in Pauli-Villars
regularization. The renormalized pressure then takes the final form
\begin{eqnarray}
p_{\rm ren}&=&
-{\cal E}_{\rm ren}+\dot\phi^2(t)
-\frac{m_0^4}{96\pi^2}-\frac{V(t)}{48\pi^2}m_0^2
-\frac{\ddot{V}(t)}{288\pi^2}\nonumber\\
&&+\int\!\frac{{\rm d^3}k}{(2\pi)^3}\,\frac{1}{2\omega_k^0}\left\{ 
\left({(\omega_k^0)}^2+\frac{\vec k^2}{3}\right)
(2{\rm Re}f_k^{\overline{(2)}}(t)+|f_k^{\overline{(1)}}(t)|^2)
\right.\nonumber\\
&&\hspace{7mm}+\left(\frac{1}{6{(\omega_k^0)}^2}
-\frac{m_0^2}{24{(\omega_k^0)}^4}\right)
\int\limits_0^t\!{\rm d}t\,
\cos{2\omega^0_k(t-t')\stackrel\dots{V}(t')}\nonumber\\
&&\hspace{7mm}+\left(\frac{1}{12{(\omega_k^0)^2}}
+\frac{m_0^2}{24{(\omega_k^0)}^4}\right)\cos{(2\omega_k^0 
t)}\ddot{V}(0)\nonumber\\
&&\hspace{7mm}+|\dot{f}_k^{\overline{(1)}}(t)|^2
-2\omega_k^0{\rm Re}(i \dot{f}_k^{\overline{(2)}*}(t) 
+if_k^{\overline{(1)}}(t)\dot{f}_k^{\overline{(1)}*}(t))
\biggr\}\; .
\end{eqnarray}
\section{Numerical analysis}  
\label{num_an}
As discussed in section 3 the computation of the fluctuation term
$\Delta {\cal M}^2(t)$ which determines the back-reaction of the
quantum fluctuations to the classical field $\phi(t)$ 
can be reduced to the computation of the
mode functions $f_k^{\overline{(1)}}(t)=f_k(t)$ and
$f_k^{\overline{(2)}}(t)$ via (\ref{numerik}). 
The differential equations satisfied by these functions,
(\ref{fdiffeq}) and (\ref{f2diffeq}), have been solved using
an improved Runge-Kutta-scheme with six steps \cite{schwarz}.
We have also checked the accuracy
of $f_k^{\overline{(2)}}(t)$ by computing it via
 the alternative expression (\ref{f2inteq}).
 
The momentum integrals are finite; however, the convergence is
not very good, as the integrand decreases only as an inverse power
of $\omega_k^0$. Furthermore the integrand is oscillating rapidly.
If we denote the upper limit of the momentum integration as
$K$ the integrals occuring in
$\Delta {\cal M}^2 $, the energy ${\cal E}$
and the pressure $p$ converge only as
$I(K)=I_\infty+C /(\omega_K^0)^2$ to their value $I_\infty$. 
We have determined the values $I_\infty$ by fitting the integrals 
as a function of $K$ to a function of this type.
This way the numerical integration can be
limited to moderate values of $K \simeq 20 \div 40$. 
A delicate part of the
momentum integration is the region of small momenta.
There the mode functions develop a parametric resonance
before their back-reaction to the classical field
$\phi(t)$ suppresses the amplitude of the latter one
and the system ``shuts off'' \cite{BAVHL}, i. e. 
reaches a stationary oscillating behaviour. We have chosen
momentum intervals varying from $\Delta k \simeq 10^{-4}$
at small momenta to $\Delta k \simeq .1$ for larger ones.
The time steps were chosen as $\Delta t \simeq 0.001$.
The quality of the numerical procedure was monitored 
by the conservation of energy which was fulfilled with
an accuracy of better than $3 \%$. With the accuracy
given above we had to compute the mode functions for $1000$
values of momentum. The computations can be performed on
a modern PC or small workstation, they take then several hours
of CPU time.

\section{Results and Conclusions}
\label{conclus}

We have presented here a computational scheme for solving
the one-loop nonlinear relaxation equations with a manifestly
covariant regularization and renormalization. 

The results are presented in Figs. 3 - 9 for two sets of 
parameters, similar to those of \cite{BVHLS}:
$\lambda/8\pi^2 = 0.1, m=1 $ (as the general unit) and $
\phi(0)= 5,~ \dot{\phi}(0)=0 $ and $ \phi(0)= 1,~
\dot{\phi}(0)=0$, respectively. In Figs.\ \ref{Fig3} and\ 
\ref{Fig8} we display the
classical amplitude $\phi(t)$, in Figs.\ \ref{Fig4} and 
\ \ref{Fig9} we show the
growth of the fluctuation integral
$\Delta {\cal M}^2 (t)$ (see (\ref{numerik})). 
Fig.\ \ref{Fig5} shows the 
absolute value of the integrand (to be integrated with the
measure $k^2dk/\omega_k^0$)
of the second fluctuation integral in (\ref{numerik})
vs. $\omega_k^0$ in a
double-logarithmic scale for the same
set of parameters and for $t=37.5 $.
 The amplitude of the
integrand is seen to fall off as $(\omega_k^0)^{-4}$.
In Fig.\ \ref{Fig6} we plot
the classical and fluctuation energies as well as the total
energy for the first parameter set. 
The pressure is displayed for the same parameter set in
Fig.\ \ref{Fig7}. Its average as $t\to \infty$ is seen to be
${\cal E}/3$ as found previously in \cite{BoHoVeSa} .
These numerical results show the same qualitative and similar 
quantitative features as those obtained in
\cite{BVHLS}.  An initial large amplitude oscillation leads 
after a short time interval - depending on the initial amplitude -
to a strong excitation of the low
momentum fluctuation modes as expected from a parametric resonance.
Then after a considerable decrease of the classical amplitude
an essentially stationary oscillating regime is reached. In contrast
to a computation in Hartree-approximation in \cite{BVHLS}
we find that even for $\phi(0)=1$ a considerable amount of 
energy is transferred to the fluctuation modes. 
The main difference
between a small and large initial value of the classical
field consists in the time that is needed for the
fluctuations to build up.

The field theory we have considered here was a rather simple
one; as discussed in \cite{BVHLS,BAVHL} one would not expect the
one-loop approximation to be a good here nor has the system
been found previously to display thermalization. Here this
simple model served as a toy example for demonstrating the method.
More interesting systems are of course gauge theories with many
light degrees of freedom. There the differential equations
for the gauge field fluctuations form coupled systems.  
We should like to stress that the method as developed in sections
\ref{pertex} and \ref{renorm}
can be easily generalized to deal with such
 coupled channel systems. This
generalization is entirely analoguous to the treatment of coupled
channel one loop computations in \cite{Baa1}.

\newpage
\section*{Figure captions}
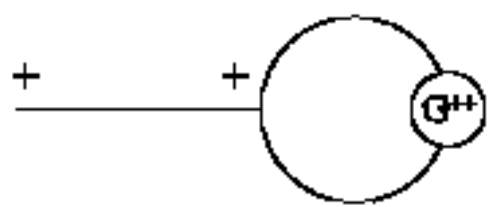
\begin{figure}
\caption{Diagrammatic representation of the 
one-loop equation (\ref{phidgl}).}
\label{Fig1}
\end{figure}
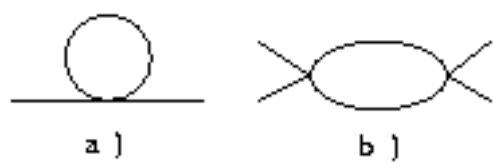
\begin{figure}
\caption{Renormalization parts: a) Tadpole diagram,
Eq. (\ref{tadpole}), b) Fish diagram,
Eq. (\ref{fish}).}
\label{Fig2}
\end{figure}
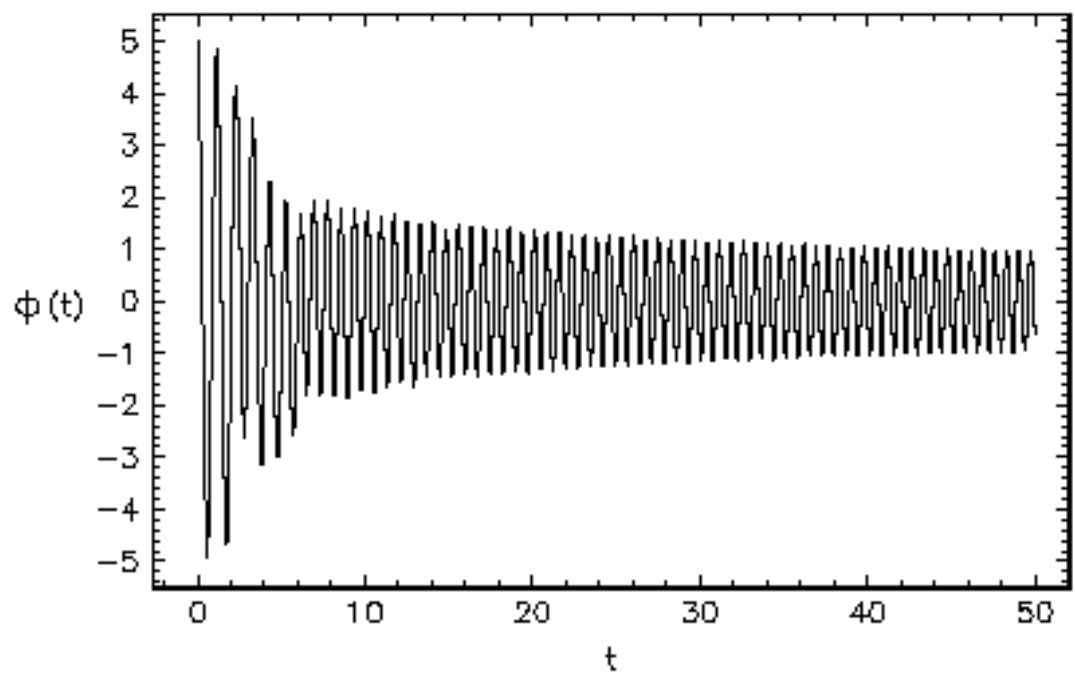
\begin{figure}
\caption{The classical amplitude $\phi(t)$ for 
$\lambda/8\pi^2 = 0.1$ and $\phi(0) = 5$.}
\label{Fig3}
\end{figure}
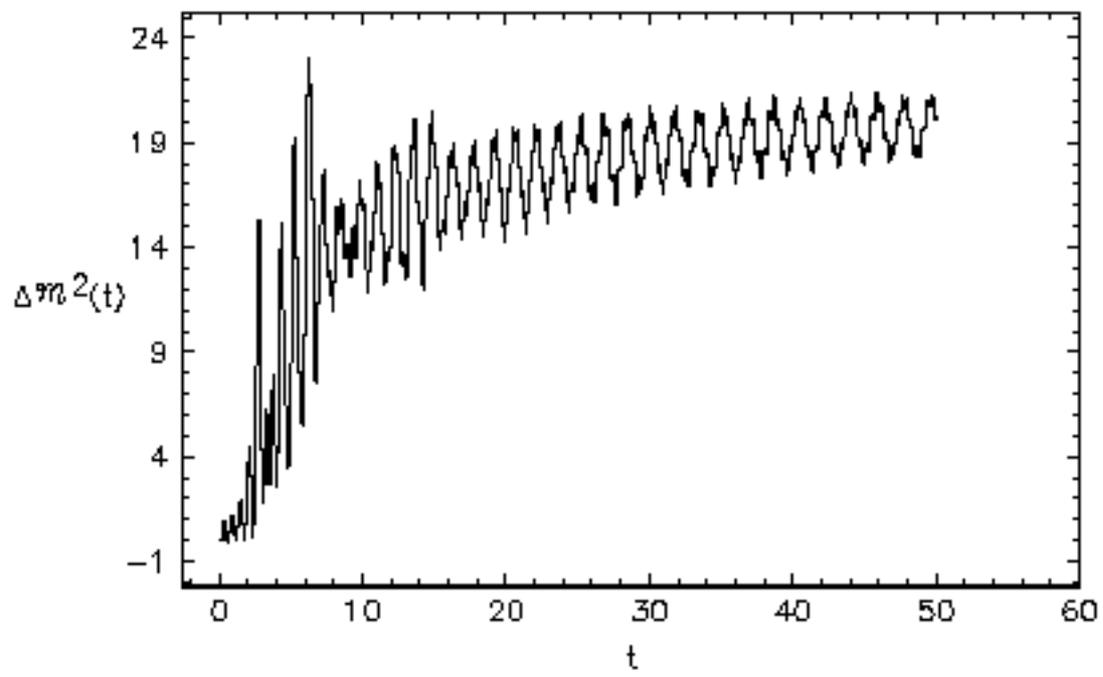
\begin{figure}
\caption{The fluctuation integral $\Delta {\cal M}^2(t)$ 
for the same set of parameters.}
\label{Fig4}
\end{figure}
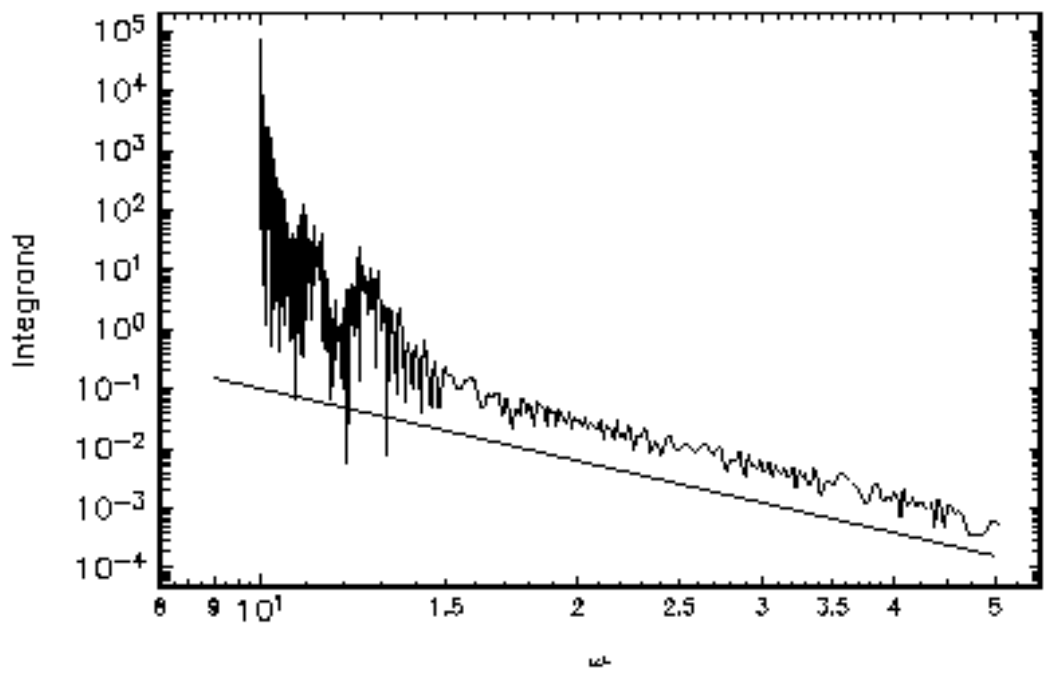
\begin{figure}
\caption{The fluctuation integrand as explained in the text.
The absolute value of the integrand is plotted vs.
$\omega_k^0$ on a double
logarithmic scale. The straight line indicates a power
behaviour as $(\omega_k^0)^{-4}$.}
\label{Fig5} 
\end{figure}
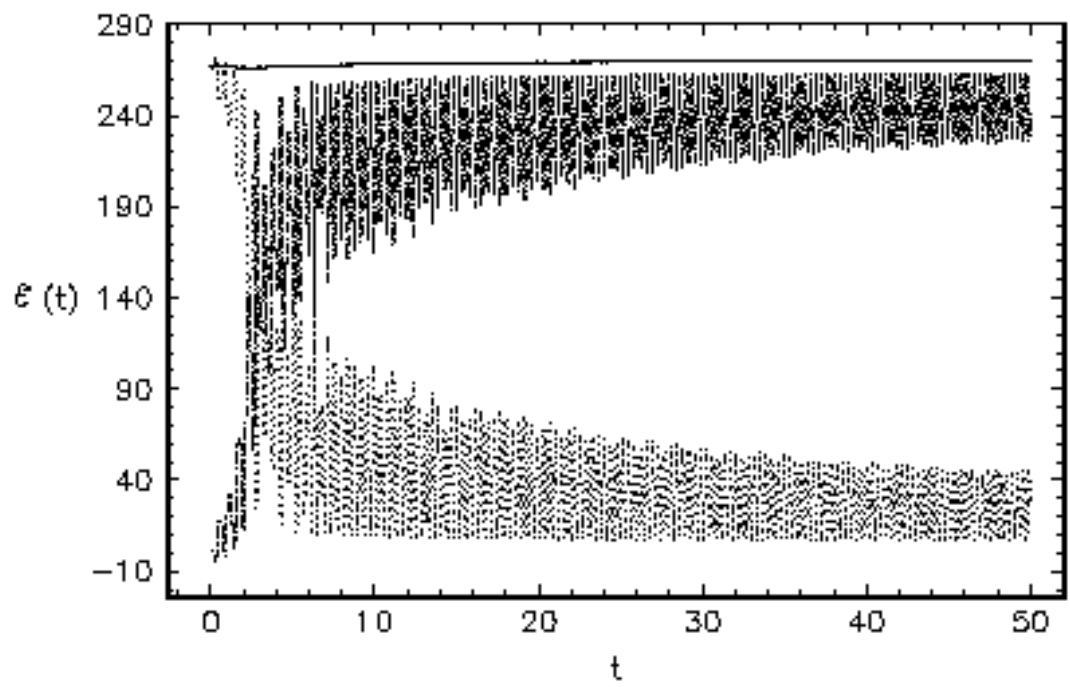
\begin{figure}
\caption{The classical and fluctuation energies as 
a function of time for the same parameters:
classical energy (short dashed line), the fluctuation energy
(long dashed line) and total energy (solid line).}
\label{Fig6} 
\end{figure}
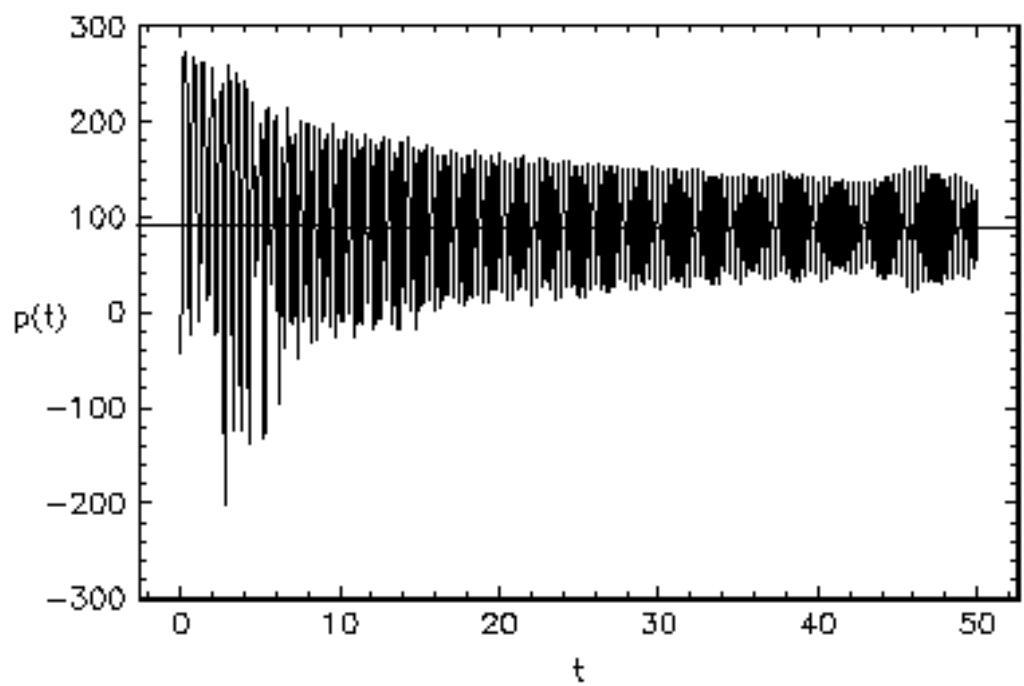
\begin{figure}
\caption{The pressure as a function of time for the same
parameter set. The horizontal line indicates the value
$p={\cal E}/3$.}
\label{Fig7}
\end{figure}
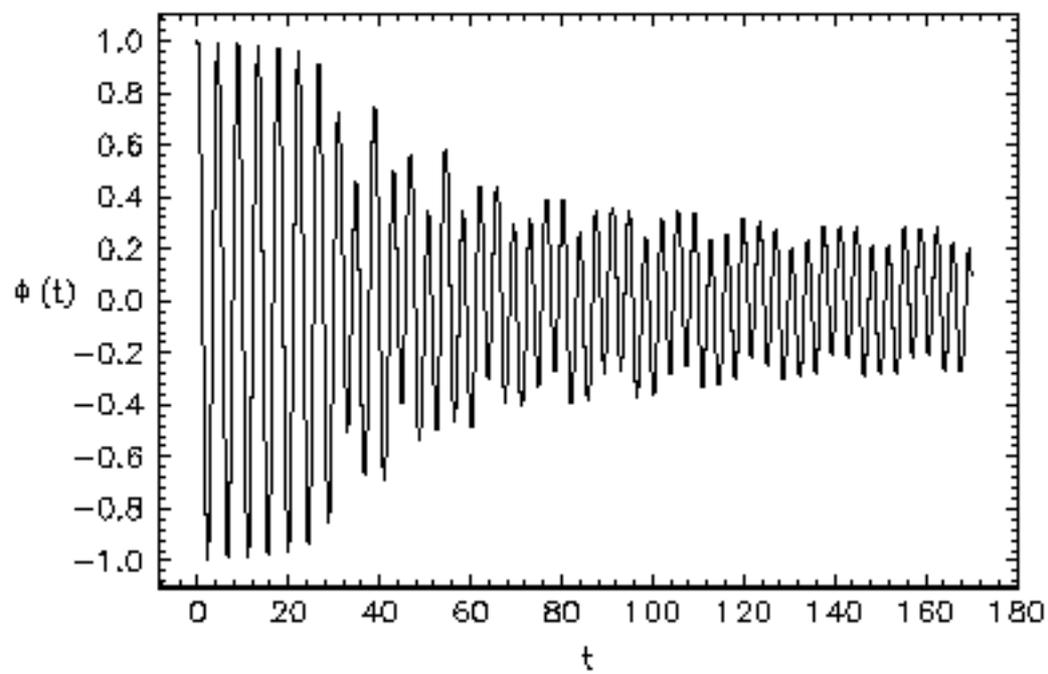
\begin{figure}
\caption{The classical amplitude $\phi(t)$ for
$\lambda/8\pi^2 = 0.1$ and $\phi(0) = 1$.}
\label{Fig8} 
\end{figure}
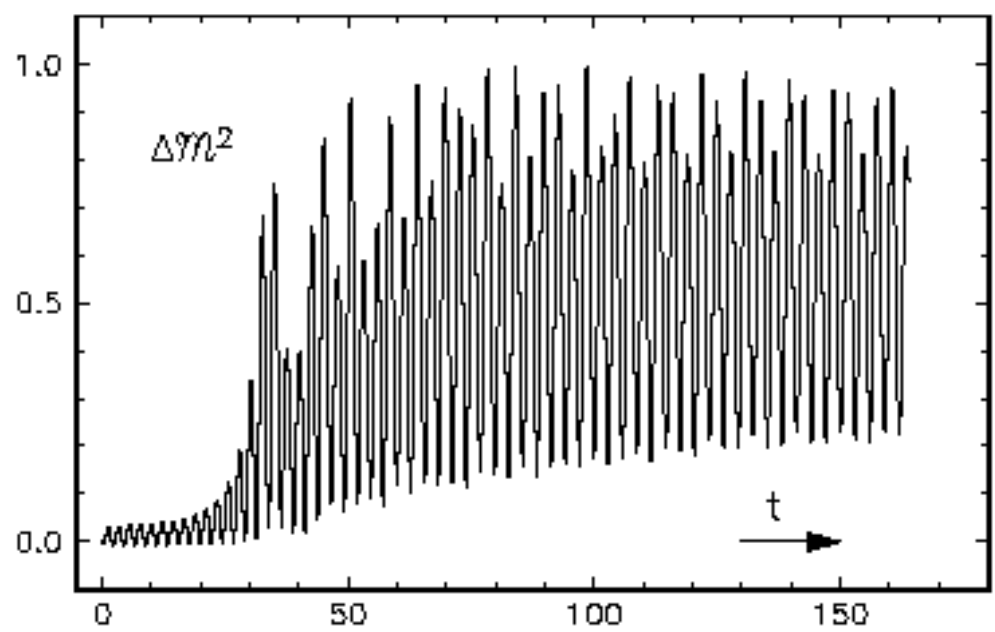
\begin{figure}
\caption{The fluctuation integral $\Delta {\cal M}^2(t)$ 
for the same set of parameters.}
\label{Fig9}
\end{figure}

\newpage
\begin{thebibliography}{99}
\bibitem{AbPi}see L. F. Abbott, S.-Y. Pi:
{\em Inflationary Cosmology}; World 
Scientific, Singapore, 1986 for a collection of the
early literature.

\bibitem{BuFoHeWa} W. Buchm\"uller, Z. Fodor, T. Helbig and
D. Walliser,  Ann. Phys. {\bf 234}, 260 (1994).

\bibitem{FaKaRuSha} K. Farakos, K. Kajantie, K. Rummukainen and
M. Shaposhnikov, Nucl. Phys. {\bf B425}, 67 (1994).

\bibitem{FaKaLaRuSha}K. Farakos, 
K. Kajantie, M. Laine, K. Rummukainen and M. Shaposhnikov,
Nucl. Phys. (Proc. Suppl.) {\bf 47}, 705 (1996).

\bibitem{Jan} K. Jansen,  Nucl. Phys. (Proc. Suppl.) 
{\bf 47}, 196 (1996).

\bibitem{Hadro} see e. g. {\em Hot Hadronic Matter}, J. Letessier, H.
H. Gutbrod and J. Rafelski, Eds., Plenum Press, New York (1995)
for a summary of recent developments.

\bibitem{AlSt}A. Albrecht,
P. J. Steinhardt, M. S. Turner and F. Wilczek,  Phys. Rev. Lett. 
{\bf 48}, 1437 (1982).

\bibitem{AbFaWi} L. F. Abbott, E. Farhi and M. Wise,
Phys. Lett. {\bf 117B}, 29 (1982).

\bibitem{BVHLS}D. Boyanovsky, H.J. de Vega, R.Holman, D.S. Lee,
A. Singh, Phys. Rev. {\bf D51}, 4419 (1995). 

\bibitem{BAVHL}D. Boyanovsky, M. D'Attanasio,
H. J. de Vega, R. Holman and D.-S. Lee, 
{\em New Aspects of Reheating}, to appear in
{\em Proceedings of the School on String Gravity and
Physics at the Planck Scale}, Erice, September 1995,
N. Sanchez, Ed., hep-ph/9511361.

\bibitem{Son} D. T. Son, {\em Classical Preheating and
Decoherence}, Univ. of Washington preprint
UW/PT-96-01, January 1996, hep-ph/9601377 ;
{\em Reheating and thermalization in a simple scalar
model}, Univ. of Washington preprint UW/PT-96-05,
hep-ph/9604340.

\bibitem{KESCM} Y. Kluger, J. Eisenberg, B. Svetitsky, F. Cooper
and E. Mottola, Phys. Rev. {\bf D48}, 190 (1993).

\bibitem{BoVeHo} D. Boyanovsky, H. J. de Vega and
R. Holman, Phys. Rev. {\bf D51}, 734 \\ (1995).

\bibitem{CKMP}F. Cooper, Y. Kluger, E. Mottola and
J. P. Paz,  Phys. Rev. {\bf D51}, 2377 (1995).

\bibitem{Baa1}J. Baacke  Z. Phys. {\bf C47}, 263 (1990);
{\em ibid. }{\bf C47}, 619 (1990); {\em ibid. }{\bf C53}, 407 (1992).

\bibitem{BaaJu}J. Baacke and S. Junker,
Phys. Rev. {\bf D49} , 2055 (1994);
{\em ibid. }{\bf D50}, 4227 (1994).

\bibitem{BaaDai}J. Baacke and T. Daiber, Phys. Rev. 
{\bf D51}, 795 (1995).

\bibitem{Baa2}J. Baacke,  Phys. Rev. {\bf D51}, 6760 (1995).

\bibitem{CalHu}E. Calzetta and B. L. Hu,  Phys. Rev.
{\bf D35}, 495 (1988); {\it ibid.} {\bf D37}, 2838 (1988).

\bibitem{We}S.~Weinberg,  Phys. Rev. {\bf D9}, 3357 (1974).

\bibitem{CaCoJa}C. G. Callan, S. Coleman and R. Jackiw, 
Ann. Phys. (NY) {\bf 59}, 42 (1970).

\bibitem{Chr}S. M. Christensen, Phys. Rev.  {\bf D14},
2490 (1976); {\em ibid.} {\bf D17}, 946 (1978).

\bibitem{BoHoVeSa} D. Boyanovsky, H. J. de Vega, R. Holman
and J. F. J. Salgado, hep-ph/9608205, to be published
in Phys. Rev. D.

\bibitem{CoHaKlMo} F. Cooper, S. Habib, Y. Kluger and
E. Mottola, Los Alamos preprint, hep-ph/9610345.

\bibitem{schwarz}H. R. Schwarz: {\em Numerische Mathematik}.
B. G. Teubner, Stuttgart 1986.
\end{thebibliography}
\end{document}